\documentclass[AMA,LATO1COL]{WileyNJD-v2}

\articletype{Article Type}%

\received{26 April 2016}
\revised{6 June 2016}
\accepted{6 June 2016}

\begin{document}

\title{Deep brain stimulation with a computational model for the cortex-thalamus-basal-ganglia system and network dynamics of neurological disorders.}

\author[1]{Hina Shaheen*}

\author[1,2]{Roderick Melnik}

\authormark{Shaheen \textsc{et al}}

\address[1]{M3AI Laboratory, MS2Discovery Interdisciplinary Research Institute, Wilfrid Laurier University,75 University Avenue West, Waterloo, Ontario, N2L 3C5, Canada}

\address[2]{BCAM - Basque Center for Applied Mathematics, 
Alameda de Mazarredo 14, E-48009 Bilbao, Spain}

\corres{*Hina Shaheen,  \\{shah8322@mylaurier.ca}}

\abstract[Summary]{Deep brain stimulation (DBS) can alleviate the movement disorders like Parkinson's disease (PD). Indeed, it is known that aberrant beta $(13-30\si{Hz})$ oscillations and the loss of dopaminergic neurons in the basal ganglia-thalamus (BGTH) and cortex characterize the akinesia symptoms of PD. However, the relevant biophysical mechanism behind this process still remains unclear. Based on the prior striatal inhibitory model, we propose an extended BGTH model incorporating medium spine neurons (MSNs) and fast-spiking interneurons (FSIs) along with the effect of DBS. We are focusing in this paper on an open-loop DBS mode, where the stimulation parameters stay constant independent of variations in the disease state, and modifications of parameters rely mainly on trial and error of medical experts. Additionally, we propose a novel combined model of the cerebellar-basal-ganglia thalamocortical network, MSNs, and FSIs, and show new results that indicate that Parkinsonian oscillations in the beta-band frequency range emerge from the dynamics of such a network. Our model predicts that DBS can be used to suppress beta oscillations in globus pallidus pars interna (GPi) neurons. This research will help our better understanding of the changes in brain activity caused by DBS, providing new insight for studying PD in the future.

}


\maketitle

\section{Introduction}\label{sec1}
Deep brain stimulation (DBS) is an effective symptomatic treatment for a range of neurodegenerative disorders such as Parkinson’s disease (PD). Although DBS has been found to alleviate motor symptoms in persons with PD, the mechanisms of action are still largely unknown, making DBS parameter selection problematic. Currently, the DBS operates predominantly in an open-loop mode, in which the stimulation characteristics stay constant regardless of disease state alterations, and parameter modifications are primarily done by trial and error by experienced clinicians \cite{lu2020erratum}. While the loss of dopaminergic neurons in the substantia nigra is known to be the etiology of PD, the impact on the basal-ganglia (BG) networks is not fully understood, despite the fact that beta-band oscillations are widely accepted. The rise in beta-band oscillations ($13-30\si{Hz}$) of the BG is a hallmark neural dynamic characteristic of PD \cite{yu2019oscillation}. These excessive synchronous oscillations will further impair the thalamus (TH) capacity to convey motion information \cite{rubin2004high}. The origin of these aberrant synchronous oscillations is being disputed due to the complex nonlinear structure of the brain \cite{ahn2016synchronized, yu2019oscillation}. In this case, computational network models can be very useful in the analysis of brain pathology and aberrant oscillations \cite{ma2017review, shaheen2021neuron, shaheen2021analysis}.

The BG system (BGS), we are interested in here, is a collection of four brain nuclei that receive data from the whole cortex and send it primarily to the frontal cortex through the thalamus. These nuclei are made up of the striatum, the subthalamic nucleus (STN), the globus pallidus pars interna (GPi), and the globus pallidus pars externa (GPe). Importantly, dopamine deficiency causes an imbalance in activation for the striatal direct and indirect pathways that are the primary channels that link the striatum to the BG's downstream structure \cite{zhai2018striatal}. The direct pathway is comprised of medium spine neurons (MSNs) in the striatum expressing dopamine D1 receptors (D1 MSNs), substantia nigra pars reticulata (SNr) and GPi. The indirect pathway is composed of MSNs expressing dopamine D2 receptors (D2 MSNs), GPe and STN. In recent years, modelling investigations of PD have mostly focused on the cortex, thalamus, and BG. Earlier, Terman and Rubin presented a computer network model of the BG and TH based on conductance \cite{rubin2004high}. So et al. \cite{so2012relative} modified this model to more precisely represent the discharge characteristics of Parkinsonian and healthy states. This model is extensively employed to investigate the mechanism and therapy of PD, despite the fact that it ignores the function of the striatum's internal network \cite{yu2019oscillation}. Numerous biophysical analyses have revealed that the striatum plays a role in PD movement \cite{zhai2018striatal,ahn2016synchronized,yu2019oscillation}, and chronic dopamine deprivation induces substantial alterations in striatal synaptic plasticity \cite{deutch2007striatal, yu2020review}. Although the striatum plays an essential role in PD, it is sometimes neglected. The striatum's interior is mostly composed of four different types of neurons, with MSNs constituting the vast majority of striatal neurons \cite{yu2019oscillation}. The striatal fast-spiking interneuron (FSI) receives excitatory synaptic effects from the cortex and thalamus, as well as inhibitory projections from other striatal neurons. D2 MSNs appear to be more excited than D1 MSNs because of their inherent excitability \cite{aosaki2010acetylcholine}. Several striatum inhibitory models have been presented. Gruber et al. \cite{gruber2003modulation} created a basic model that included essential ionic currents. Humphries et al. developed a novel dopamine-modulated multicompartment model based on Izhikevich's canonical spiking model \cite{humphries2009dopamine}. The model by  McCarthy et al. solely addressed the function of M-current in MSNs \cite{mccarthy2011striatal}. In addition, a novel MSN model comprising major ion channels to examine the involvement of the striatal inhibitory microcirculation and the BGTC model including the striatum microcirculation has also been investigated \cite{yu2019oscillation}.

In terms of neurophysiological alterations, the pathological changes that occur in the human brain as a result of PD and other movement disorders may be well understood. However, one distinguishing feature of the PD disease is that an oscillatory activity has been observed in brain signals obtained from clinical recordings \cite{beudel2019linking,lofredi2019beta} or animal models of movement disorders \cite{deffains2019parkinsonism}. While the occurrence of such pathological oscillations has been widely established, the genesis and spread of these brain activities remain poorly understood.
To improve this understanding, the use of population models can shed a lot of light on the mechanisms behind pathological neural activities in PD, as well as on the mechanisms underlying DBS \cite{yousif2020population}. Yousif et al. utilized the Wilson-Cowan approach \cite{wilson1972excitatory} to simulate populations of excitatory and inhibitory neurons linked into the brain network. 
\begin{figure}[h]
\centering
\includegraphics[scale=0.35]{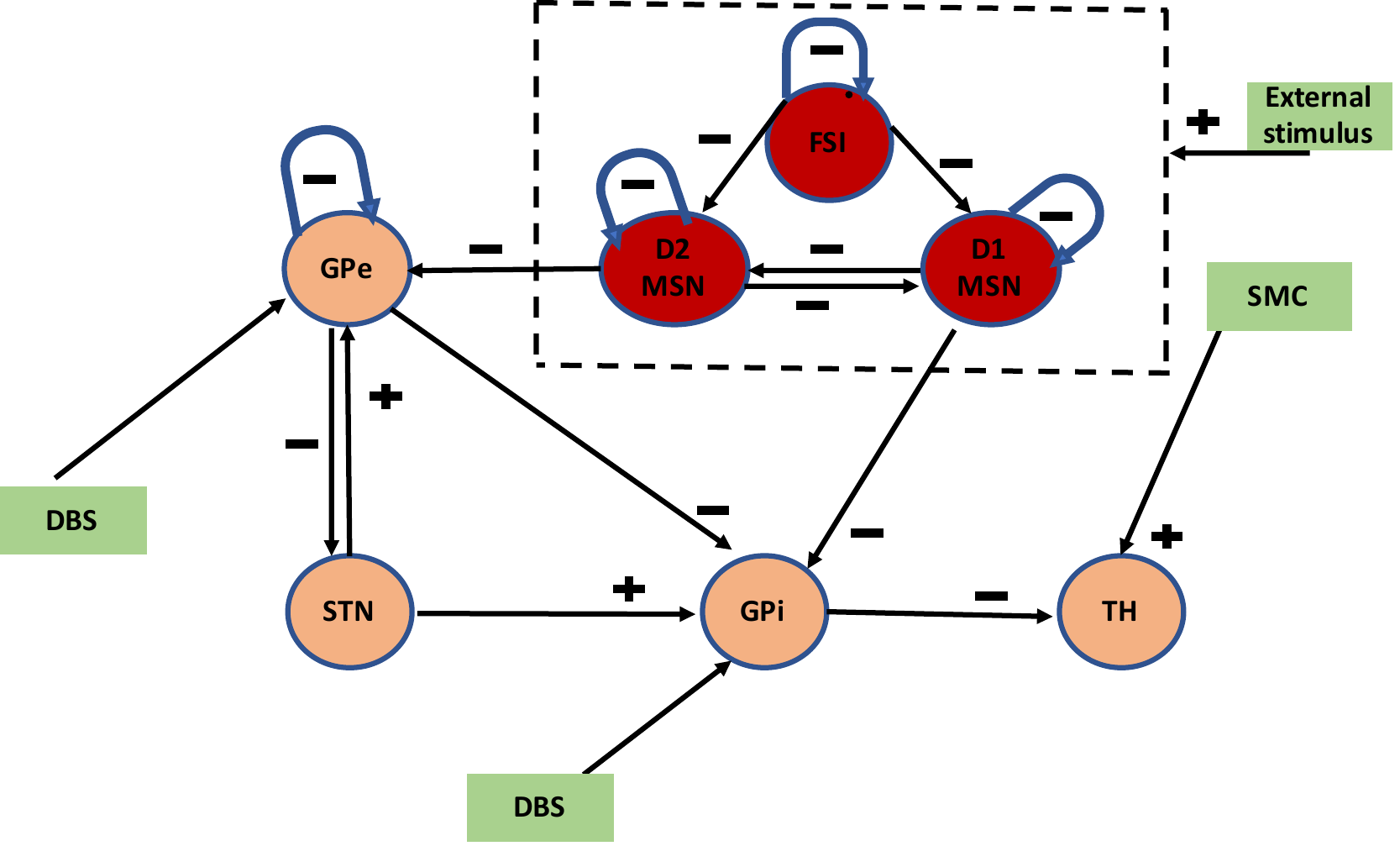}
\caption{(Color online) The basal ganglia-thalamus network mode connectivity schematic diagram.
The globus pallidus pars externa (GPe), subthalamic nucleus (STN), globus pallidus pars interna (GPi), D1 medium spine neurons (D1 MSNs), D2 medium spine neurons (D2 MSNs), fast-spiking interneurons (FSIs), thalamus (TH), and sensorimotor cortex  (SMC) are the key components of the model. Each nucleus in the striatum gets external current input. The symbol ‘$+$' denotes the excitatory current, and ‘$-$' denotes the inhibitory current.}
 \label{fig:1}   % Give a unique label
\end{figure}

\begin{figure}[h]
\centering
\includegraphics[scale=0.35]{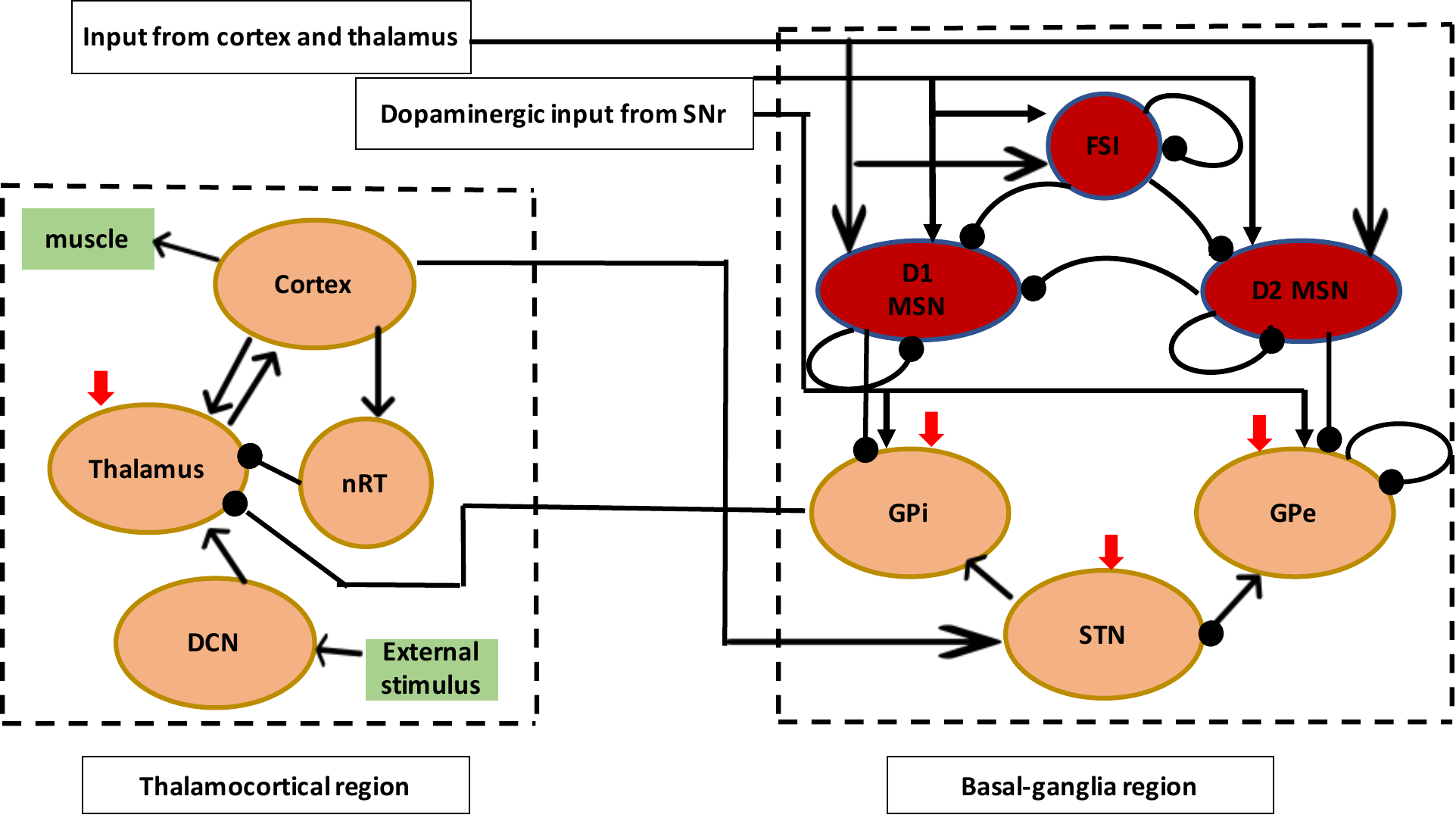}
\caption{(Color online) The cerebellar-basal-ganglia thalamocortical network. In this case, we integrate the two networks by adding MSNs and FSIs. Excitatory connections are shown by arrows, dopaminergic are shown by solid arrowheads, whereas inhibitory connections are denoted by spherical arrowheads. The four brain areas targeted by deep brain stimulation (DBS) are shown by solid red arrows.}
 \label{fig:2}   % Give a unique label
\end{figure}

To proceed with the model development which would better account for the above challenges in the context of a specific group of subcortical nuclei, we note that the DBS is a neurosurgical procedure in which high-frequency, brief monopolar pulse trains are supplied via an implanted pulse generator and injected into widely used portions of the BG network \cite{wong2020comprehensive}. In the present study, we investigate an improved computational model of the basal ganglia-thalamus (BGTH) network based on the model originally proposed by Rubin and Terman \cite{rubin2004high} for the effects of DBS on the evolution of PD, incorporating four brain nuclei. Specifically, the Hodgkin-Huxley (HH) neurons are employed in this model to replicate the four basic nuclei in BGTH   containing MSNs and FSIs and to further explore possible scenarios of PD progression and treatment. In this approach, four major nuclei (STN, GPe, GPi, and TH) are linked together by excitatory and inhibitory synaptic connections to create the BG network, which reacts to SMC input. In Fig.~\ref{fig:1}, the BGTH  model is schematized, where it is seen that striatum microcirculation is utilized to study the effect of DBS in PD. In addition, we propose a novel integrated model of the cerebellar-basal-ganglia thalamocortical network, as well as MSNs and FSIs, and provide new findings demonstrating that Parkinsonian oscillations in the beta-band are caused by the dynamics of such a network. We employ a population representation of the combined thalamocortical basal-ganglia network model with MSNs and FSIs as depicted in Fig.~\ref{fig:2}. The cortical-basal ganglia-thalamo-cortical loop was separated into two pathways: a direct pathway (cortex-striatum-GPi-thalamus) that initiated and facilitated voluntary movement, and an indirect pathway (cortex-striatum-GPe-STN-GPi-thalamus) that inhibited movement.
D1-receptor-expressing striatal MSNs largely project to the direct pathway, whereas D2-receptor-expressing MSNs primarily project to the indirect pathway. Dopaminergic input from the substantia nigra pars reticulata (SNr) to the striatum enhances direct route activity via D1 receptors and lowers indirect pathway activity via D2 receptors, enabling movement. The scale of the cerebellar-basal-ganglia thalamo cortical network model is validated from the experimental study\cite{herrington2016mechanisms}. This model is based on the Wilson-Cowan approach. We present the firing pattern and spectral characteristics of neurons in simulated healthy and Parkinsonian states. Increased inhibition of MSNs by the striatum generates oscillatory activity in the beta-band of the BGTH , which leads to bursting oscillations of the Gpi neurons, as seen via the local field potential. Furthermore, dopamine deprivation can produce beta-band oscillations in the GPi.

The rest of the paper is organized as follows. In Section 2, we describe our model in its different components: (i) basal-ganglia network model, (ii) cerebellar-basal-ganglia thalamocortical network, (iii) deep brain stimulation dynamics. In Section 3, we present numerical simulations based on the developed computational model for the cortex-thalamus-basal-ganglia system and network dynamics of neurological disorder. Finally, we discuss our results and outline future directions in Section 4.

\section{Methods}\label{sec2}
\subsection{Mathematical modelling of the basal ganglia-thalamus network model}
The core BGTH  model, taken here as a starting point, was initially presented by Yu et al. \cite{yu2019oscillation} which was modified from the original Rubin-Terman model \cite{rubin2004high} to more closely reflect experimental evidence on neuron firing characteristics and was used to examine the impact of neuronal activation on relay reliability through the TH. In this model, four main nuclei (STN, GPe, GPi, and TH) and FSI, as well as two kinds of MSNs, are linked together by excitatory and inhibitory synaptic connections to create the BGTH  network. The schematic representation of the model is given in Fig.~\ref{fig:1}. The sensorimotor cortex (SMC) input denotes the
excitatory input from the SMC to the TH. For simplicity, we suppose that each nucleus has 10 neurons and that the number of each nucleus is equal. We add MSNs and FSIs using the same connection approach to the prior sparse connection framework model developed by Yu et al \cite{yu2019oscillation}. Similar to the previous modelling works, the single compartment conductance-based Hodgkin–Huxley (HH) models modified by So et al.\cite{so2012relative} are used to describe the dynamical behaviours of GPe, GPi, STN, and TH neurons with the only distinctions being in parameter settings and changes in neuron-specific currents. In the equations that follow, variables pertaining to STN, GPe, GPi, and TH neurons are indicated by the subscript or superscript  Sn, Ge, Gi and Th, respectively. The predominant ionic currents of these four neurons
 contain the leak current $I_L$, the potassium current 
$I_K$, the sodium current $I_{Na}$, the low-threshold T-type
$Ca^{2+}$ current $I_T$, the high-threshold $Ca^{2+}$ current $I_{Ca}$
and the $Ca^{2+}$-activated after-hyperpolarization $K^+$ current
$I_{AHP}$. The membrane dynamics of STN, GPe, GPi, and TH neurons
can then be described by the following governing equations:

\begin{equation}
     \frac{dV_{Th}}{dt}=\frac{1}{c_m}\bigg(-I_{Na}-I_K-I_L-I_T-I_{Gi\rightarrow{Th}}+I_{SMC}\bigg),
\end{equation}
\begin{equation}
     \frac{dV_{Sn}}{dt}=\frac{1}{c_m}\bigg(-I_{Na}-I_K-I_L-I_T-I_{Ca}-I_{AHP}-I_{Gi\rightarrow{Sn}}+I_{Snapp}\bigg),
\end{equation}

\begin{equation}
     \frac{dV_{Ge}}{dt}=\frac{1}{c_m}\bigg(-I_{Na}-I_K-I_L-I_T-I_{Ca}-I_{AHP}-I_{Sn\rightarrow{Ge}}-I_{Ge\rightarrow{Ge}}-I_{D2\rightarrow{Ge}}+I_{Geapp}\bigg),
\end{equation}

\begin{equation}
     \frac{dV_{Gi}}{dt}=\frac{1}{c_m}\bigg(-I_{Na}-I_K-I_L-I_T-I_{Ca}-I_{AHP}-I_{Sn\rightarrow{Gi}}-I_{Ge\rightarrow{Gi}}-I_{D1\rightarrow{Gi}}+I_{Giapp}\bigg),
\end{equation}
where the $I_{SMC}$ represents the cortical sensorimotor signal received by the thalamus, which is specified by a sequence of random monophasic current pulses. The applied constant current is denoted by $I_{app}$. Also, following \cite{lu2020erratum, lu2019effect}, we have $I_{Snapp} = 23$, $I_{Geapp} = 7$, and $I_{Giapp} = 15$ for the pathological state. The particular formulas and relevant parameters for these neurons can be found in Table~\ref{tab:1} adopted from previous works \cite{so2012relative, yu2019oscillation}. The HH model is also used to model D1 and D2 MSNs. The potassium current, the sodium spike-producing current, the leak current, the inward-rectifying potassium (Kir) current, the L-type $Ca^{2+}$ (CaL1.3) current, and the M-current are all taken into account in the previous MSN modelling studies \cite{yu2019oscillation,mccarthy2011striatal}, with the M-current being regulated by cholinergic neurons in the striatum. Importantly, Kir and CaL1.3 channels have been found to influence the biophysically unique characteristics of MSNs in the previous studies \cite{aosaki2010acetylcholine,olson2005g}. The membrane potential of MSNs reads as: \cite{yu2019oscillation}:
\begin{equation}
     \frac{dV_{D1/D2}}{dt}=\frac{1}{c_m}\bigg(-I_{Na}-I_K-I_L-I_M-I_{CaL}-I_{Kir}-I_{Syn}+I_{D1/D2ext}\bigg),
\end{equation}
 where $c_m=1\si{\mu Fcm^{-2}}$ is the membrane capacitance, $E$ is the reverse potential of each current, respectively, and $g$ is the membrane conductance. The corresponding currents are $I_K=g_Kn^4(V-E_K), I_{Na}=g_{Na}m^3h(V-E_{Na}), I_{L}=g_{L}(V-E_L), I_{M}=g_{M}M(V-E_M)$ and $I_{Kir}=g_{Kir}a_{\infty}(V)(V-E_{Kir})$. The activation and inactivation gating variables are defined as follows \cite{yu2019oscillation}:
 \begin{equation}
     \frac{dx}{dt}=\frac{x_{\infty}-x}{\tau_x},
\end{equation}
where $x=(m,h,n,M)$, $x_{\infty}$ is the steady state function, and $\tau_x$ is the decay time given by:
\begin{equation}
     x_{\infty}=\frac{\alpha_x}{\alpha_x+\beta_x},
\end{equation}
where the relevent parameters are found in Table~\ref{tab:1}.
The voltage-dependent activation variable, the inactivation variable, the calcium-dependent inactivation variable, and factor $F$ comprising membrane biophysical characteristics are the major components of the basic model for L-type $Ca^{2+}$ current \cite{tuckwell2012quantitative}. In the study of striatal dynamics, the Goldman–Hodgkin–Katz form is frequently used for the modelling of $F$ \cite{wolf2005nmda}. The L-type $Ca^{2+}$ current in the present model used earlier by Wolf et al.\cite{wolf2005nmda} and it is described as:

\begin{equation}
     I_{CaL}=\overline{P}_{Ca}\frac{(z_{Ca}F)^2}{RT}\cdot\frac{[Ca]_0-[Ca]_iexp(-Vz_{Ca}F/RT)}{1-exp(-Vz_{Ca}F/RT)},
\end{equation}
where the specific parameter values are found in \cite{wolf2005nmda,yu2019oscillation}.

Moreover, as in \cite{yu2019oscillation}, we set $I_{D1ext}=6$, $I_{D2ext}=5.5$ and the synaptic currents are represented by $I_{Syn}$. For D1 MSNs, $I_{Syn}$ comprises inhibitory projections from D2 MSNs, D1MSNs, and FSIs, i.e. $I_{D2\rightarrow D1}$, $I_{D1\rightarrow D1}$, and $I_{FSI\rightarrow D1}$. For D2 MSNs, it contains $I_{D1\rightarrow D2}$, $I_{D2\rightarrow D2}$, and $I_{FSI\rightarrow D2}$. To date, to describe the electrophysiological characteristics of FSIs, a number of modelling approaches for interneurons have been developed \cite{nomura2003synchrony,planert2010dynamics,yu2019oscillation,yu2020review}. The firing equilibrium between MSNs is influenced by FSIs in the striatum. The FSI model is adopted from \cite{yu2019oscillation}, originally proposed by \cite{nomura2003synchrony}, and is given as:
\begin{equation}
     \frac{dV_{FSI}}{dt}=\frac{1}{c_m}\bigg(-I_{Na}-I_K-I_L-I_{FSI\rightarrow{FSI}}-I_{gap}+I_{ext}\bigg),
\end{equation}
where $I_{ext}=5$ is the external constant current (this chosen value has been motivated by \cite{yu2019oscillation}), $I_{FSI\rightarrow{FSI}}$ is the synaptic connection between FSIs, $I_{gap}=g_{gap}(V-V_{pre})$ is the gap junction, respectively, and all other specific formulas and related parameters can be found in the previous work \cite{nomura2003synchrony,yu2019oscillation}. In a similar fashion, $I_{a\rightarrow{b}}$ represents the synaptic current
from neuron $a$ to neuron $b$ adopted from \cite{so2012relative,yu2019oscillation,lu2020erratum}. It can be described as:
\begin{equation}
     I_{a\rightarrow{b}}=g_{a\rightarrow{b}}(V_b-V_{a\rightarrow{b}})\sum_ks_a^k,
\end{equation}
where $g_{a\rightarrow{b}}$ is the synaptic conductance, $V_{a\rightarrow{b}}$ represents the synaptic reversal potential and the parameters in these formulas follow those described in detail in the works \cite{yu2019oscillation,so2012relative}.

\subsection{Mathematical modelling of the cerebellar-basal-ganglia thalamocortical network}
In the present Section, we propose a new numerical model of the combined thalamocortical basal-ganglia network with FSI and two kinds of MSNs as shown in Fig.~\ref{fig:2}. This model, based on the previous studies \cite{yousif2017network,yousif2020population}, includes a thalamocortical part and a basal-ganglia part. The thalamocortical part of the model contains a cortical population, a cerebellar population, and two thalamic populations \cite{yousif2017network}. The basal-ganglia part includes an STN population, a population representing the GPe, and mainly a population representing the GPi, as this brain region is also implicated in movement disorders, particularly the pathology associated with PD. In order to couple these two networks, we incorporate important connections reported in the literature, such as a cortical drive to the STN, with information transmitted via the hyperdirect route. In addition, the GPi is considered the basal-ganglia's output and provides inhibitory output to the thalamus. As in earlier studies \cite{yousif2017network,yousif2020population}, the network gets ascending drive to the thalamus via the cerebellar population. Moreover, here we include MSNs in the striatum expressing D1 MSNs, D2 MSNs and FSIs. Each FSI blocks the three D1 MSNs and three D2 MSNs closest to it. It also transmits inhibitions to two FSIs. According to the likelihood of mutual projections in the previous models \cite{bahuguna2015existence,ponzi2013optimal,yu2019oscillation}, we assume that each D1 MSN inhibits two D2 MSNs, two D1 MSNs, and three adjacent GPi neurons. Each D2 MSN inhibits two D2 MSNs, three D1 MSNs, and four neighbouring GPe neurons.

The model is developed by combining the Wilson-Cowan approach \cite{wilson1972excitatory} and Yu et al. model \cite{yu2019oscillation}. The Wilson-Cowan approach has been frequently used to simulate populations of excitatory and inhibitory neurons linked into networks and Yu et al. model is the BGTH  model containing
MSNs and  FSIs. The framework is based on the premise that neurons within a population are in close physical proximity to MSNs and FSI, therefore it ignores spatial interactions and solely represents temporal dynamics. As depicted in Fig.~\ref{fig:2}, we have an excitatory cortical population, two thalamic populations, the excitatory ventralis intermediate nucleus (Vim) and the inhibitory reticular nucleus (nRT), an excitatory population of cerebellar neurons indicating the deep cerebellar nuclei (DCN), the cerebellum's main output, an excitatory population representing the STN, two inhibitory populations representing the GPe and the GPi, FSI and D1, D2 MSNs. Therefore,
the model is composed of seven first-order coupled differential
equations and variables corresponding to neurons of the STN, GPe, GPi, DCN, nRT, TH and cortex are represented by the subscript Sn, Ge, Gi, Dn, Rt, Th and Cx, respectively, as summarized below:

\begin{equation}
     \frac{dE_{Cx}}{dt}=\frac{1}{\tau_{Cx}}\bigg(-E_{Cx}+(k_e-E_{Cx})\cdot \mathcal{H}_e ({w_1E_{Th}})\bigg),
\end{equation}

\begin{equation}
     \frac{dE_{Th}}{dt}=\frac{1}{\tau_{Th}}\bigg(-E_{Th}+(k_e-E_{Th})\cdot \mathcal{H}_e ({w_2E_{Cx}}-w_3I_{Rt}+w_4E_{Dn}-w_5I_{Gi})\bigg),
\end{equation}

\begin{equation}
     \frac{dI_{Rt}}{dt}=\frac{1}{\tau_{Rt}}\bigg(-I_{Rt}+(k_i-I_{Rt})\cdot \mathcal{H}_e ({w_6E_{Cx}})\bigg),
\end{equation}

\begin{equation}
     \frac{dE_{Dn}}{dt}=\frac{1}{\tau_{Dn}}\bigg(-E_{Dn}+(k_e-E_{Dn})\cdot \mathcal{H}_e (ext)\bigg),
\end{equation}

\begin{equation}
     \frac{dI_{Ge}}{dt}=\frac{1}{\tau_{Ge}}\bigg(-I_{Ge}+(k_i-I_{Ge})\cdot \mathcal{H}_i ({w_7E_{Sn}}-w_8I_{Ge}-I_{D2\rightarrow{Ge}})\bigg),
\end{equation}

\begin{equation}
     \frac{dI_{Gi}}{dt}=\frac{1}{\tau_{Gi}}\bigg(-I_{Gi}+(k_i-I_{Gi})\cdot \mathcal{H}_i ({w_9E_{Sn}}-I_{D1\rightarrow{Gi}})\bigg),
\end{equation}

\begin{equation}
     \frac{dE_{Sn}}{dt}=\frac{1}{\tau_{Sn}}\bigg(-E_{Sn}+(k_e-E_{Sn})\cdot \mathcal{H}_e ({w_{10}E_{Cx}}-w_{11}I_{Ge})\bigg),
\end{equation}
where, $E_i$ $(i=Cx, Th, Dn, Sn)$ and $I_j$ $(j=Rt, Ge, Gi)$ reflect the number of activated neurons in the appropriate excitatory or inhibitory population at a particular time \cite{yousif2020population}. The two functions $\mathcal{H}_i(x)$ and $\mathcal{H}_e(x)$ are monotonically increasing sigmoid
functions. Each of these functions represents the fraction of firing cells in a population at a particular level of average membrane potential activity $x(t)$ as shown here:
\begin{equation}
  \mathcal{H}_p(x)=\frac{1}{1+exp(-b_p(x-\theta_p))}-\frac{1}{1+exp(b_p\theta_p)},
\end{equation}
where $p=e,i$, $b_p$ and $\theta_p$ are constants \cite{yousif2020population}, $x$ is the
level of input activity, and $\tau$ is the time constant of the change in the fraction of non-refractory cells firing in a population in response to the change in the cells' average membrane potential activity over time. Therefore, to add the essence of striatum input the membrane potential of MSNs and FSIs has been added directly to GPe and GPi populations at a particular level of x(t). Finally, $w_n$ $(n= 1, \dots 11)$ shows the strength of connections between two populations. It reflects the average number of contacts per cell multiplied by the average postsynaptic current produced in the postsynaptic cell by a presynaptic action potential \cite{yousif2020population}. It is noteworthy that the Dn population in Eq. 14, is not affected by the dynamics of the other populations and simply serves as an input to the Vim population \cite{yu2019oscillation}. As a result, in this model, the Dn population tends to have a stable value and does not fluctuate. The network dynamics for FSI and D1, D2 MSNs are the same as presented in Eqs. 5-10 in Section 2.1. All other relevant parameters are given in Table~\ref{tab:2}\cite{wilson1972excitatory, yousif2020population}.
%\subsection{Tables}
\begin{table}
\caption{Parameter set for the BGTH  network.}
\centering
\label{tab:1}       % Give a unique label
%
% Follow this input for your own table layout
%
\begin{tabular}{p{3cm}p{3cm}p{3cm}p{3cm}}
\hline\noalign{\smallskip}
Parameter & Value & Parameter & Value \\
\hline\noalign{\smallskip}
$E_K$  &$-100 (mV)$  &$E_L$  &$-67 (mV)$ \\
$E_{Na}$  &$50 (mV)$  &$E_M$  &$0 (mV)$\\
$g_K$  &$80 mS/cm^2$  &$g_L$  &$0.1 mS/cm^2$ \\ 
$g_M$  &$2 mS/cm^2$  &$g_{Na}$  &$100 mS/cm^2$ \\
$E_{Kir}$  &$-82 mV$  &$g_{KirD1}$  &$0.175 mS/cm^2$\\
 $g_{KirD2}$  &$0.14 mS/cm^2$  &$Q$ &$3.209$\\
  $\tau_x$  &$\frac{1}{\alpha_x +\beta_x}$  &
$\alpha_{m}$& $\frac{(0.32(V+54))}{(1-exp(-(V+54)/4))}$\\
$\beta_{m}$& $\frac{(0.28(V+27))}{(-1+exp(-(V+27)/4))}$ &$\alpha_{h}$ & $0.1*exp(-(V+50)/18))$\\
$\beta_{h}$& $\frac{4}{(1+exp(-(V+27)/5))}$ &$\alpha_{n}$& $\frac{0.032*exp((V+52)}{(1-exp(-(V+54)/4))}$\\
$\beta_{n}$&$0.5*exp(-(V+57)/40))$ & $\alpha_{M}$& $\frac{Q*10^{-4}((V+30)}{(1-exp(-(V+30)/9))}$\\
$\beta_{M}$& $-\frac{Q*10^{-4}((V+30)}{(1-exp(-(V+30)/9))}$ & $a_{\infty} (V)$&$\frac{1}{(1+exp((V-(-102))/13))}$\\
\end{tabular}
\end{table}
\FloatBarrier
\subsection{DBS modelling}

Importantly, previous studies have shown that open-loop DBS-STN stimulation can boost Th relay reliability \cite{chen2014simulating, dorval2010deep, dorval2009deep}. However, the GPi is also considered as a therapeutically useful target for stimulation \cite{lu2020erratum}. In addition, the inhibitory waveform from the GPi and  GPe to the TH has a significant impact on relay reliability. Several additional works \cite{ahn2016synchronized,cakir2021computational,} reported that thalamic relay reliability was a significant variable to evaluate the therapeutic efficiency of DBS. Therefore, certain closed-loop approaches evaluated on HH type neurons are provided to restore thalamic relay reliability in order to enhance DBS. Thus, directly manipulating the GPi and GPe inhibitory waveforms by stimulation of the GPi and GPe improves the predictability of the effect of deep brain stimulation on relay reliability \cite{liu2016closed}. The GPi, on the other hand, has a bigger structure than the STN, which can improve treatment safety and make targeting easier \cite{lu2020erratum}. In the present study, GPi and GPe are chosen as the target for stimulation for the model presented in Section 2.1. The DBS is modeled as \cite{lu2020erratum}: 
\begin{equation}
     I_{DBS}=i_D H(\sin (2 \pi t/ \rho_D ))\cdot [1-H(\sin (2 \pi(t+\delta_D)/ \rho_D ))],
\end{equation}
where $i_D=200\si{\mu A/cm^2}$ denotes the stimulation amplitude, chosen here for numerical experiments, $\delta_D =0.6\si{ms}$ is the duration of each impulse, $f$ is
the frequency and $\rho_D =6\si{ms}$ is the stimulation period. The DBS current is added in to the model presented in Section 2.1 to the membrane potential equations
of each GPi and GPe neurons directly, for example,
the GPi equation is modified as follows: 
\begin{equation}
     \frac{dV_{Gi}}{dt}=\frac{1}{c_m}\bigg(-I_{Na}-I_K-I_L-I_T-I_{Ca}-I_{AHP}-I_{Sn\rightarrow{Gi}}-I_{Ge\rightarrow{Gi}}-I_{D1\rightarrow{Gi}}+I_{Giapp}+I_{DBS}\bigg),
\end{equation}

However, we simulated DBS of the network population model presented in Section 2.2 via the application of a high-frequency input to the GPi in the presence of FSI and D1/D2 MSNs. As previously stated, this modification adds a new term to the equation for the stimulated population; for example, the GPi equation will now have the form:
\begin{equation}
     \frac{dV_{Gi}}{dt}=\frac{1}{\tau_{Gi}}\bigg(-I_{Gi}+(k_i-I_{Gi})\cdot \mathcal{H}_i ({w_9V_{Sn}}-I_{D1\rightarrow{Gi}}+I_{DBS})\bigg),
\end{equation}
where the relevant parameters are found in Table~\ref{tab:1} and Table~\ref{tab:2} \cite{damodaran2015desynchronization,lu2020erratum}.

%\subsection{Tables}
\begin{table}
\caption{Parameter set for thelmocortical basal-ganglia network.}
\centering
\label{tab:2}       % Give a unique label
%
% Follow this input for your own table layout
%
\begin{tabular}{p{3cm}p{3cm}p{3cm}p{3cm}}
\hline\noalign{\smallskip}
Connection & Parameter & Healthy state  & Parkinsonian state \\
\hline\noalign{\smallskip}
$Th\rightarrow Cx$  &$w_1$  &$20$  &$20$ \\
$Cx  \rightarrow Th$ &$w_2$& $5$& $5$\\
$Rt \rightarrow Th$ &$w_3$& $8$ &$8$\\
$Dn\rightarrow Th$ &$w_4$ & $25$ &$20$\\
$Gi \rightarrow Th$ &$w_5$ &$15$ &$15$\\
$Cx \rightarrow Rt$& $w_6$ &$5$ &$5$\\
$Sn \rightarrow Ge$& $w_7$ & $19$ &$5$\\
$Ge \rightarrow Ge$ &$w_8$ & $5$ &$5$\\
$Sn\rightarrow Gi$& $w_9$ &$15$ &$15$\\
$Cx \rightarrow Sn$ &$w_{10}$ &$20$ &$20$\\
$Ge \rightarrow Sn$ &$w_{11}$ &$20$ &$20$\\
$ext \rightarrow Dn$ & $ext$ & $3.42$ & $3.42$\\
\end{tabular}
\end{table}

\section{Results}\label{sec3}

\subsection{Basal-ganglia network model}

  \begin{figure}[h]
\centering
\includegraphics[scale=0.6]{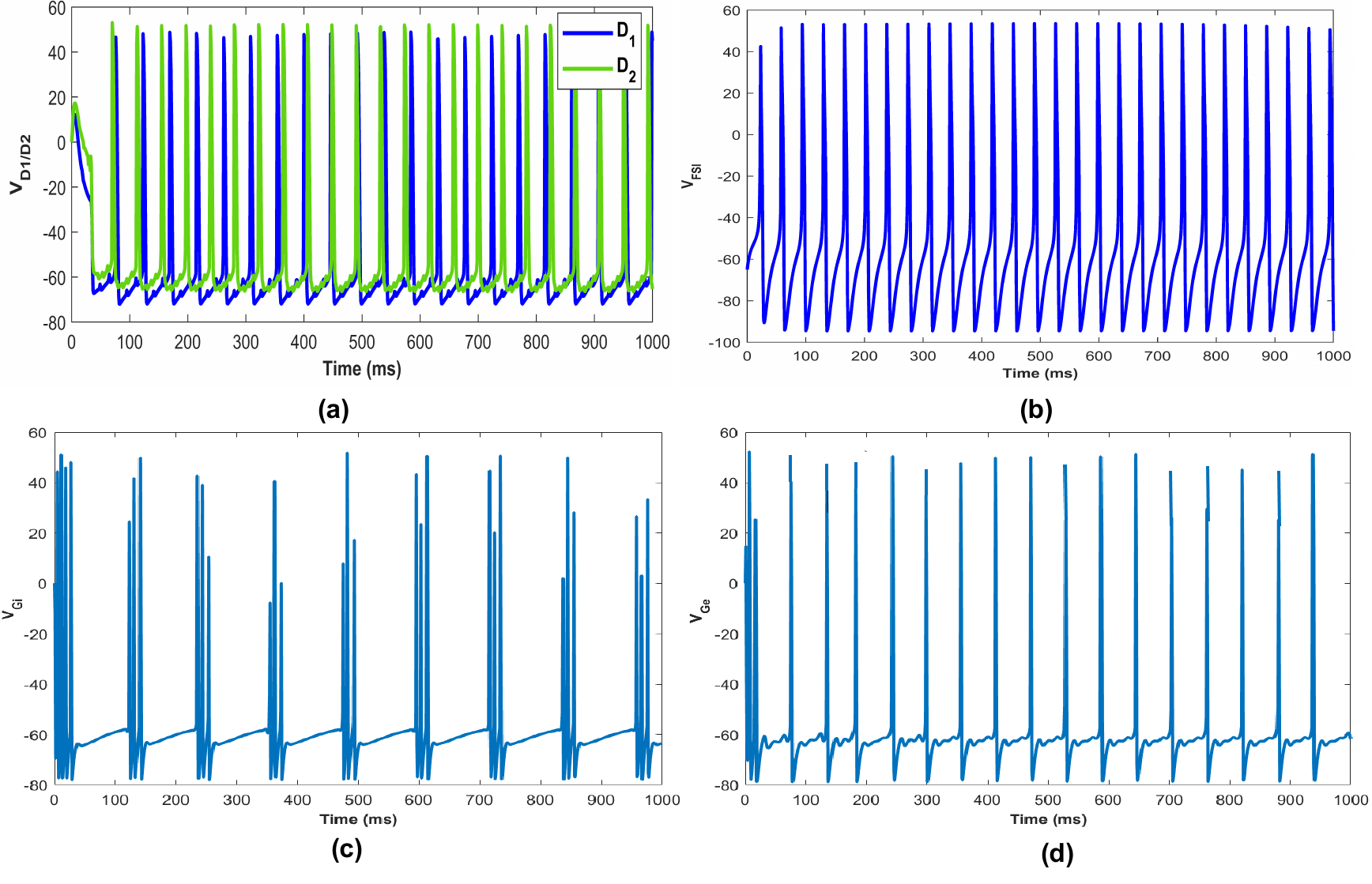}
\caption{(Color online) Membrane voltages ($\si{mV}$) of the (a) D1/D2 MSNs, (b) FSIs, (c) GPi neurons and (d) GPe neurons.}
 \label{fig:3}   % Give a unique label
\end{figure} 
In this Section, a numerical study has been performed for quantifying the effects of DBS, MSNs and FSI on the BGTH network. The MATLAB platform has been used for all network simulations. Fig.~\ref{fig:3} depicts the firing characteristics of a single neuron under the aforementioned model parameters presented in Section 2.1. The FSI response is seen in Fig.~\ref{fig:3}a, and it consists of a continuous high-frequency sequence with a brief action potential. Fig.~\ref{fig:3}b. shows the activation of D1 and D2 MSNs. D2 MSN has a greater discharge rate than D1 MSN. The presented results are consistent with experimental evidence for these dynamic behaviours \cite{gertler2008dichotomous}. Fig.~\ref{fig:3}c-d exhibits the firing of GPi and GPe neurons, respectively. The GPi and GPe neurons are involved in the control of autonomic movement, and the major characteristic of the Parkinsonian condition is enhanced by synchronous activity between neurons in a burst-like manner \cite{lu2020erratum}. For simplicity, we have illustrated the results of the GPi and GPe neurons out of four neurons as presented in Section 2.1. The reason for this is because the striatum, as the major input structure of the BGTH , plays a key role in controlling the dynamic activity of neurons such as GPe and GPi. Notably, dopamine deficiency alters the internal structure of the striatum. Meanwhile, the equilibrium between direct and indirect channels has been disrupted, which is strongly connected to the BGTH  Parkinsonian state.  
  \begin{figure}[h]
\centering
\includegraphics[scale=0.6]{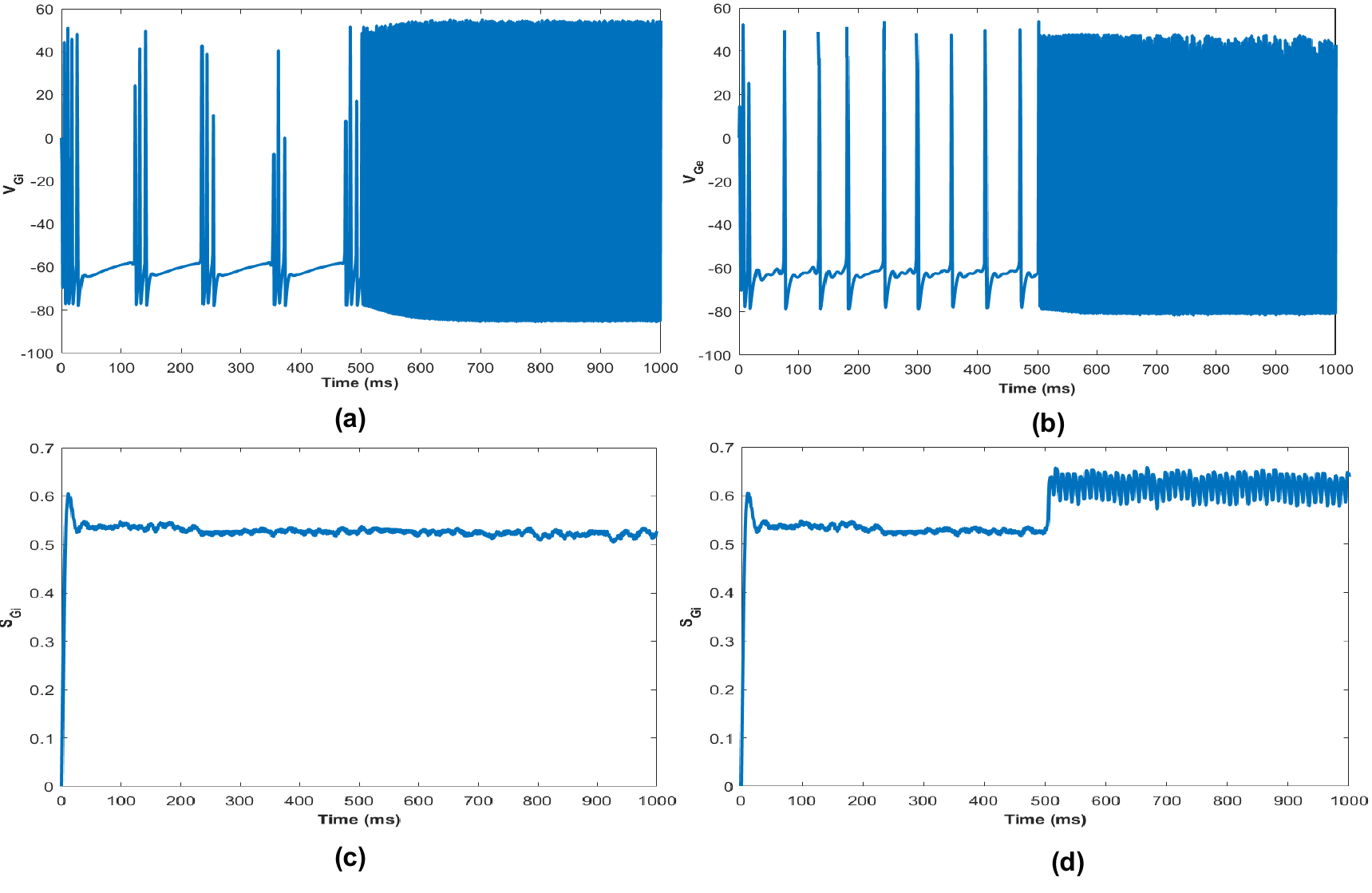}
\caption{(Color online) Effects of open-loop DBS on the (a) GPi neurons, (b) GPe neurons. (c) Synaptic variable of GPi neurons ($S_{Gi}$). (d) Effects of open-loop DBS on the $S_{Gi}$.}
 \label{fig:4}   % Give a unique label
\end{figure} 
 
Moreover in our computational experiments, DBS was applied to the BGTH  network when $t=500\si{ms}$, via a square pulse as described above in Section 2.3. With DBS input given by $120\si{Hz}$ to the GPe and GPi neurons, Fig.~\ref{fig:4}a-b shows that the GPe and GPi neurons exhibit bursting behaviour and synchronization. It is noted that after $t=500\si{ms}$ the bursting patterns of GPi and GPe neurons are replaced by tonic spiking. In this scenario, GPi's synchronous and burst output in the presence of MSNs and FSI is potent enough to impact thalamic activity, as evidenced by the dynamics of synaptic variable $S_{Gi}$ in Fig.~\ref{fig:4}c. It can be seen that the synaptic variable $S_{Gi}$ oscillates predictably and repetitively without DBS and DBS effectively inhibits the rhythmic oscillation of the synaptic variable as shown in Fig.~\ref{fig:4}d. From Fig.~\ref{fig:4}, it is revealed that, when DBS  input was delivered to GPe and GPi, the bursting behaviour occurs relative to the PD state. DBS electrodes are generally permanently implanted in the BGTH , and the stimulator sends electrical impulses constantly and without being dependent on feedback (open-loop stimulation). The open-loop pattern might have negative consequences for patients due to the possibility of overstimulation. As a result, a closed-loop DBS may be required, in which stimulation settings are automatically changed in response to changes in current neurophysiological data.
\FloatBarrier
\subsection{Cerebellar-basal-ganglia thalamocortical network model}
In Section 2.2, we constructed a mathematical model of the thalamocortical cerebellar-basal-ganglia network of MSNs and FSI. We based our simulation on two prior studies that looked at the thalamocortical network and the subthalamic nucleus (STN)-GPe network independently \cite{yousif2020population, cakir2021computational}. The dopamine-modulated type neuron model, which consists of fast-spiking interneurons, D1 and D2 type dopamine expressing medium spiny neurons, has been used to mimic the striatal area. The conventional thalamocortical basal ganglia network neuron model, on the other hand, has been employed in the modelling of extrastriatal basal ganglia regions in which globus pallidus subregion neurons contain dopamine-dependent properties. The thalamic part of the network is mass modelled, having inhibitory input from the basal ganglia \cite{cakir2021computational}. We have explored the synchronisation of basal ganglia neuron populations in relation to the loss of synaptic connections and dopamine levels. Modeling the dopamine-dependent neuron characteristics incorporates the influence of dopamine in GP. Now, based on the developed model, we will demonstrate that such networks are capable of generating robust beta oscillations. To understand the dynamics of this cerebellar-basal-ganglia thalamocortical network, we have first analyzed the network activity as a whole. Specifically, we have simulated the network and investigated the connection parameters to identify the areas that exhibited oscillatory activities in the frequency ranges we were most interested in. That is, healthy oscillations ($>30 \si{Hz}$)\cite{beudel2015tremor,fischer2017subthalamic} and Parkinsonian beta-band oscillations ($13-30\si{Hz}$) \cite{yousif2020population}.

  \begin{figure}[h]
\centering
\includegraphics[scale=0.6]{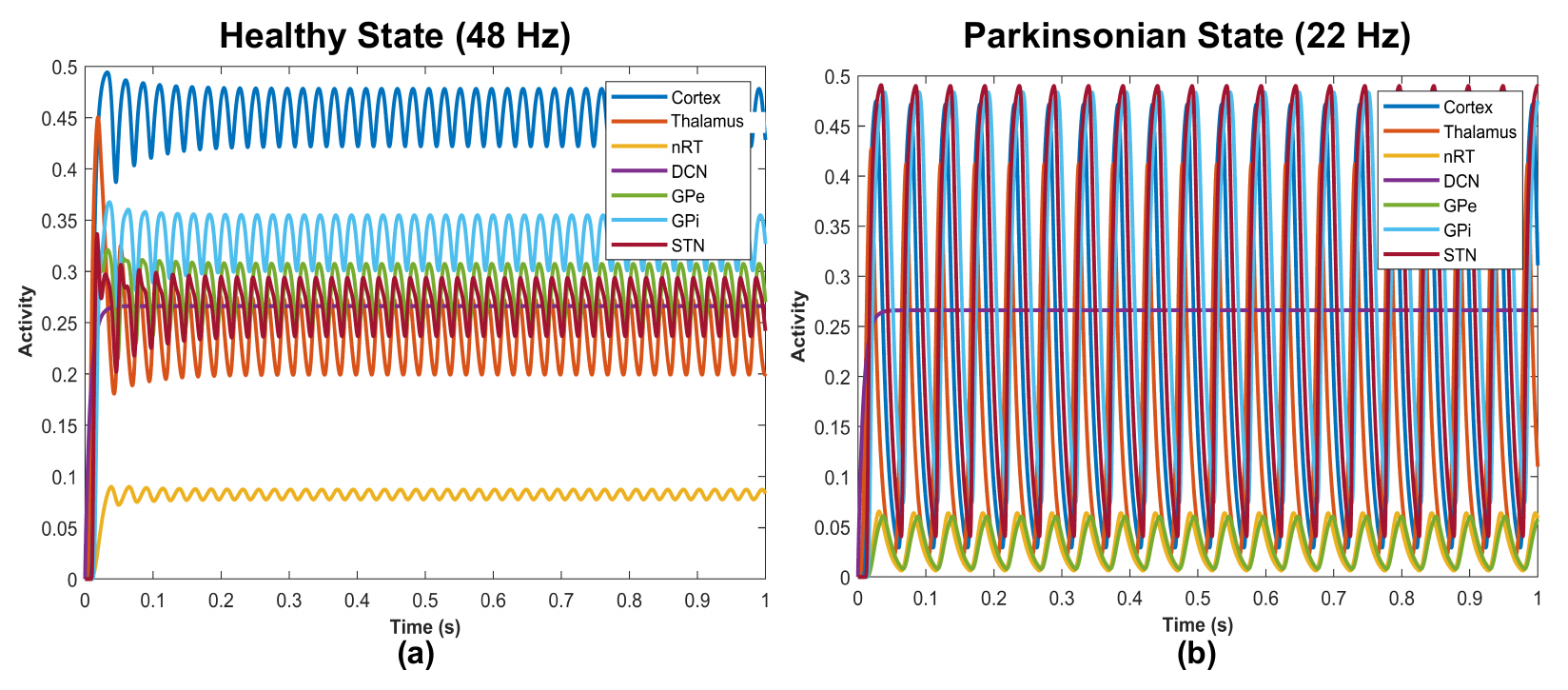}
\caption{(Color online) Activities of all populations with a selected set of parameters given in Table~\ref{tab:2}.}
 \label{fig:5}   % Give a unique label
\end{figure} 

Overall, we have discovered that the network oscillated easily throughout a substantial portion of the parameter space. With a selected set of parameters adopted from \cite{yousif2020population}, we set up the network that displayed healthy oscillations at $48 \si{Hz}$. At $20-24\si{Hz}$, the network exhibited readily Parkinsonian beta-band oscillations with the different weight parameters presented in Table ~\ref{tab:2} \cite{yousif2020population}. The changes involved in the set of parameters for beta-band oscillations are decreasing the DCN drive to the thalamus ($w_4$) and decreasing the STN to GPe connection weight ($w_7$), while the network oscillated easily in the range of $20-24\si{Hz}$. Fig.~\ref{fig:5} represents a simulation that results in oscillatory network activities for  healthy and Parkinsonian states. The network is capable of oscillating across the parameter space of interest as shown in Fig.~\ref{fig:5}a. It can be seen that in Fig.~\ref{fig:5}b, except for the DCN, all populations in the neural network oscillate at $22\si{Hz}$. Surprisingly, the amplitude of the oscillations, which indicates the fraction of neurons firing in a population, is larger in most Parkinsonian states than in a healthy state (Fig.~\ref{fig:5}a-b). Importantly, the amplitude of oscillations in the Parkinsonian state differed by population, with the STN population having the highest, followed by GPi, cortex, Thalamus, nRT, and GPe having the lowest. The thalamus, on the other hand, led the oscillations, followed by cortex, nRT, STN, GPe, and GPi in the end. Note that Fig.~\ref{fig:5} shows the results without D1/D2, MSNs synaptic projections and, as a consequence, the results for the beta-band pairwise variations are consistent with the earlier studies \cite{yousif2020population}.
\begin{figure}[h]
\centering
\includegraphics[scale=0.65]{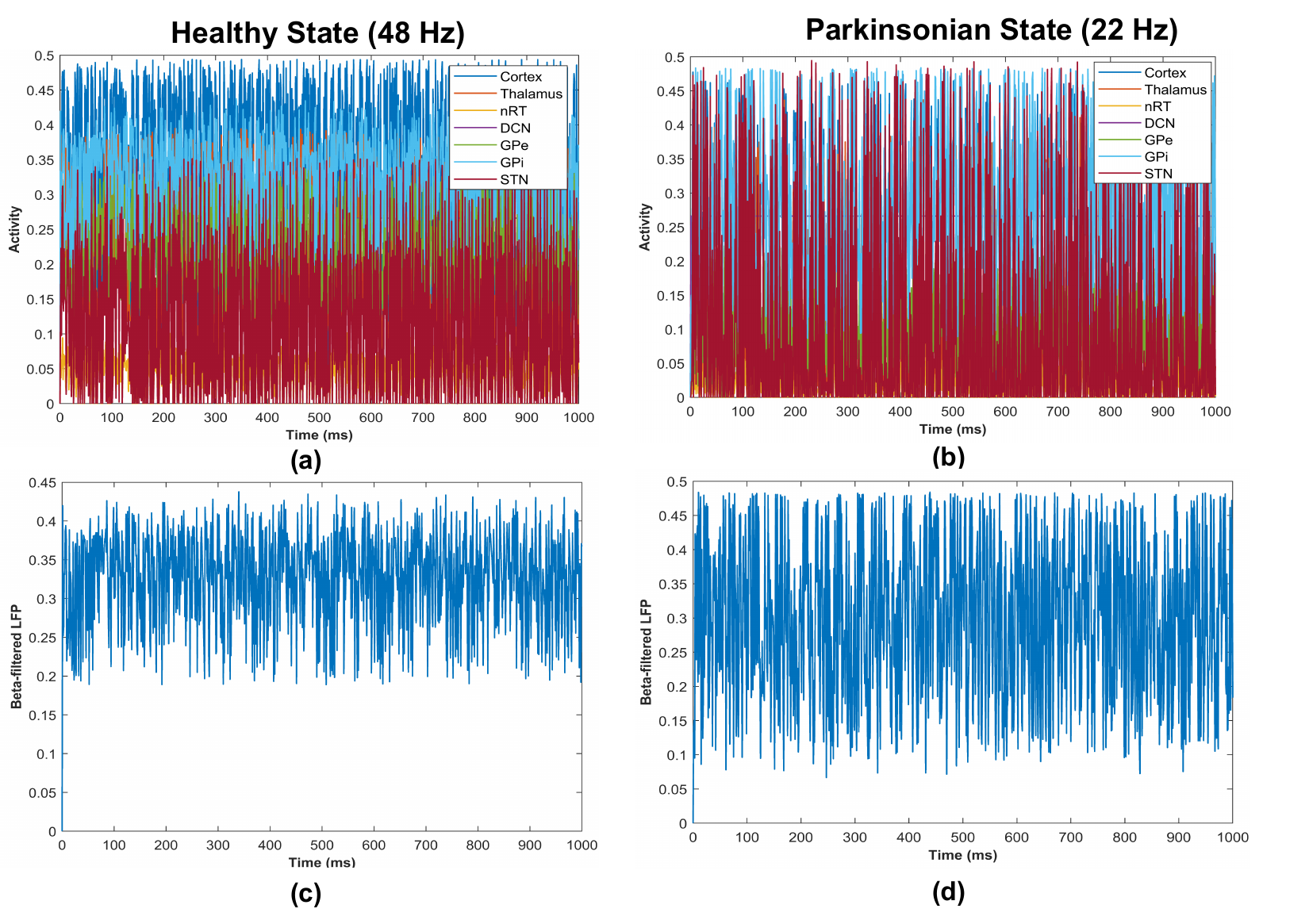}
\caption{(Color online) Firing patterns and beta-filtered LFP in the normal and Parkinsonian states. (a) Firing pattern of all populations in the presence of MSNs and FSIs, (c) beta-filtered LFP
of the GPi neurons in the normal states. (b) Firing pattern of all populations in the presence of MSNs and FSIs, (d) beta-filtered LFP of the GPi neurons in the Parkinsonian states.}
 \label{fig:6}   % Give a unique label
\end{figure}

To compare the differences between the simulated
normal and Parkinsonian states, we present the discharge pattern of a healthy state compared with the beta state in the presence of MSNs ans FSIs. These MSNs ans FSIs are projected to GPi and Gpe neurons as shown in Fig.~\ref{fig:6}. To determine whether the activity of each population changed, we examine the behaviour of each population in the normal and PD networks. In the normal state,
the bursting behaviour occurs and oscillations tend to
change non-monotonically at a peak of $48\si{Hz}$. In contrast to the rises and falls of the oscillations in the non-Parkinsonian state, the oscillations in the activity of each population remain permanent \cite{mccarthy2011striatal} as shown in Fig.~\ref{fig:6}a-b. Individual populations exhibit more pronounced subthreshold frequency oscillations due to an increase in the number of MSNs participating in the population beta rhythm \cite{yousif2020population}. This would results in a significant change in the firing pattern of the STN population followed by GPi, cortex, thalamus, nRT, and GPe. The STN shows the lowest amplitude and the cortex leads the oscillations (Fig.~\ref{fig:6}a). D2 MSNs, on the other hand, can impact GPi via D1 MSNs, including D2–D1–GPi, D2–D1–D2–GPe–STN–GPi, and D2–D1–D2–GPe–GPi pathways. By inhibiting the synaptic connection, several modulation mechanisms for D2 MSN to GPi have been explored \cite{yu2019oscillation,humphries2009dopamine}. It can be seen in Fig.~\ref{fig:6}b that GPi still exhibits oscillatory activity in the beta-band, but it is not the same as in a healthy condition. In Fig.~\ref{fig:6}b, the amplitude of 
oscillations indicates that cortex, nRT, GPe, and GPi also have significant beta-band oscillations and the firing rates of the STN and thalamus have
increased. It is seen from (Fig.~\ref{fig:6}a-b) that as the striatal output rises, the firing patterns of GPe and GPi neurons shift from random discharge to burst discharge, affecting the relay capacity of cortex, nRT and thalamus neurons. The local field potential (LFP) is more suited for quantifying clinical symptoms, and its oscillatory activity in the beta-band has been linked to Parkinson's disease symptoms \cite{deffains2019parkinsonism}. Under physiological circumstances, synaptic activity is the most significant generator of extracellular currents. LFP is strongly connected to changes in synaptic activity, and many studies have used synaptic activity to represent the properties of LFP signals indirectly \cite{herreras2016local, yu2020review}. However, LFP, on the other hand, is linked to GPi neuron synaptic dynamics, and the LFP of synchronised neurons exhibits substantial high amplitude oscillations. To simulate the LFP, the averaged synaptic activity of the GPi neurons is utilised as given in \cite{yu2019oscillation}. Additionally, the LFP of GPi after filtering through the beta-band has been presented in Fig.~\ref{fig:6}c-d. It is observed in Fig.~\ref{fig:6}d that the beta-filtered LFP of GPi was considerably enhanced as compared to the normal condition shown in Fig.~\ref{fig:6}c. The LFP power peaks at a higher frequency of $22\si{Hz}$, and the peak power is higher than in the non-Parkinsonian condition (Fig.~\ref{fig:6}c). Interestingly, however, the beta-filtered LFP also showed the bursting behaviour and a high amplification is seen in the GPe and STN cases. Compared with the normal state, changing the synaptic strength of GPi showed an increase in beta
frequency activity in the Parkinsonian state (Fig.~\ref{fig:6}a-b). This would result in a considerable rise in the LFP's amplitude, accompanied by an increase in the synchronism of the neurons. These findings showed that the GPi exhibited substantial beta activity in the PD dynamics.
\begin{figure}[h]
\centering
\includegraphics[scale=0.65]{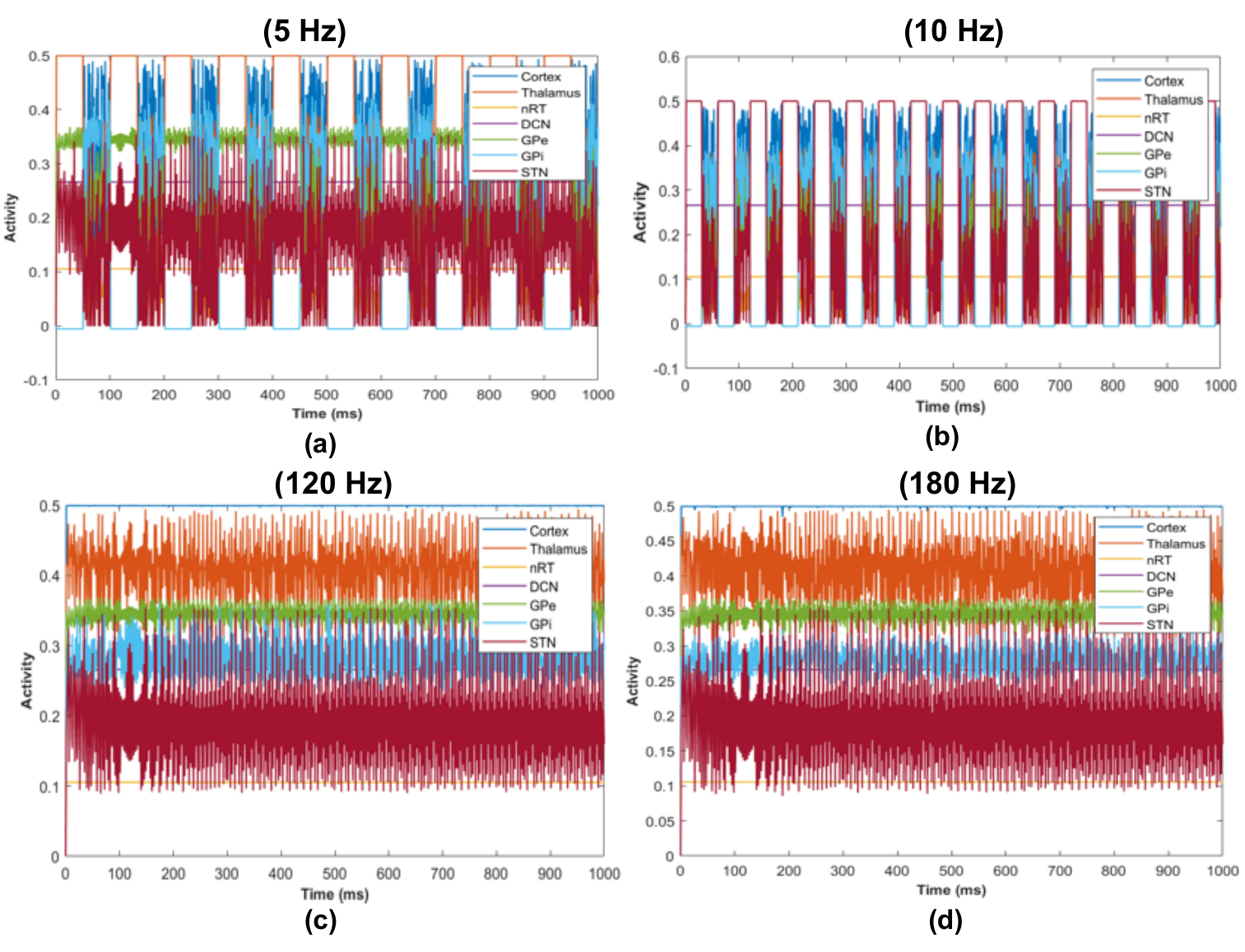}
\caption{(Color online) Impact of GPi DBS on the beta-band activity. At low frequencies, there is a decrease in the frequency of the oscillations, with the beta activity appearing to occur in bursts. This impact is visible from $120\si{Hz}$ onwards.
It is essential to notice that the network activity is oscillatory for frequencies larger than $120 \si{Hz}$, and all populations, except the DCN, oscillate at the stimulation frequency.}
 \label{fig:7}   % Give a unique label
\end{figure}
To investigate the influence of DBS on GPi at different frequencies, we simulated the application of DBS as described by Eq. 21, where the amplitude and the pulse width were chosen constants from the range $3-180\si{Hz}$ in the MSNs. According to the Fig.~\ref{fig:7}, when DBS was applied to the GPi, the network began to display beta-band activity. There is no change in the amplitude of the network activity at low frequencies. In this scenario, there is a drop in the frequency of the oscillations, with the activity appearing to occur in bursts. However, the low-amplitude high-frequency activity takes over above $50\si{Hz}$, replacing the high-amplitude low-frequency activity. It can be seen from Fig.~\ref{fig:7} that when DBS with $80\si{Hz}$ input was delivered to GPi in the presence of MSNs and FSIs, the beta frequency activity of GPi neurons was not inhabited but it was enhanced. However, when DBS with $120\si{Hz}$ input was delivered to GPi, MSNs beta oscillations were effectively inhibited relative
to the PD state. GPi beta activity was substantially diminished as the frequency of DBS was raised to $180\si{Hz}$. We discovered that DBS had an inhibitory impact on GPi beta activity when the frequency was greater than $120\si{Hz}$, and that this effect is even more pronounced as the frequency was increased. When the frequency was lower than $120\si{Hz}$, the beta activity of GPi was increased. As a consequence, we may conclude that therapeutically high frequency DBS administered to GPi is the only way to stop beta oscillations in GPi-MSNs-FSIs.

We also applied the DBS to the thalamic population, as well as to the STN and GPe populations, and found a similar trend as shown in Fig.~\ref{fig:7} for both thalamic and STN DBS simulations. The beta-band oscillations in the GPe population were suppressed once more by DBS of the GPe, while the network continued to cycle at $20\si{Hz}$. Surprisingly, at low DBS frequencies, the GPe displayed the same bursting activity as in the STN, thalamus, and GPi cases previously observed in the literature \cite{yousif2017network,yousif2020population, yu2020review}.

\section{Discussions and Conclusions}\label{sec5}
In terms of neurophysiology, the degenerative changes that occur in the human brain as a result of PD and other movement disorders may be well known. For example, it is generally known that dopamine neurons in the substantia nigra are lost in Parkinson's disease \cite{dauer2003parkinson, yousif2020population, lu2020erratum}.
On the other hand, the role of striatal inhibitory networks in modulating striatum balance has recently been studied \cite{yu2020review}. Although a better understanding of the striatum's impacts on the complete thalamic-basal-ganglia system has been highly desirable, it still faces serious challenges. In our attempts to address this issue, we have created a striatum network model using MSNs and FSIs, which we subsequently integrated into the BGTH  model. In the presence of MSNs and FSIs, we carried out DBS simulations in an open-loop fashion. In this manner, we presented an integrated mathematical model for the basal-ganglia system containing MSNs and FSIs and then embedded it into the DBS model. Our model predicts that high-frequency DBS of the GPe and GPi in the striatum may have negative consequences for patients due to possible overstimulation. As a result, closed-loop DBS procedures may be required, in which stimulation parameters are automatically modified in response to changes in ongoing neurophysiological inputs.

However, one overarching characteristic of Parkinson's disease is the presence of oscillatory activity in brain signals observed from clinical recordings \cite{beudel2019linking, lofredi2019beta} or animal models of movement disorders \cite{deffains2019parkinsonism,yousif2020population}. While the occurrence of such pathological oscillations is widely documented, the processes causing and propagating such brain activity remain unclear. To shed more light on this issue, we developed a population-level neuronal model of the basal-ganglia thalamocortical network, containing MSNs and FSIs and, more importantly, we analyzed the influence of DBS on this abnormal activity. Based on this model, we demonstrated that the FSI input can dominate the relatively simple, noisy structure of the whole model's spike train groups, supporting views that the sparse FSIs play a substantial role in the striatum \cite{tepper2004gabaergic}. They appeared to obfuscate or desynchronize the MSN-only network's structure. Our model revealed that this network was capable of displaying a wide variety of oscillatory activities up to $48\si{Hz}$, particularly in the frequency ranges associated with movement problems. As pointed out in \cite{yousif2020population}, this large range is fascinating in and of itself, since the network may represent both steady-state non-oscillatory activities and varied frequencies of pathological and healthy oscillations. It is worth mentioning that the transitions to different frequency bands were enabled by a single set of settings that resulted in a healthy condition. Our model supports the idea that enhanced GPe and GPi synchronisation leads to a greater beta activity in the loop between GPe, GPi and striatum, and rhythmicity in MSNs can contribute to pathological GPe and GPi synchrony \cite{lu2019effect,cakir2021computational}. As a result, we propose that by interrupting the rhythmicity in the MSNs, aberrant beta oscillations in GPe and GPi can be inhibited as well. 

In conclusion, we have demonstrated that a single basal ganglia-thalamus network model and the cerebellar-basal-ganglia thalamocortical network may accommodate numerous types of oscillatory activities, which have been suggested to represent both healthy brain states and pathological neural activities. By evaluating the LFP signal of GPi neurons, we have confirmed that the synaptic connection strength between MSNs and GP not only impacts the level of synchronization of GPi neurons but also results in beta-band oscillations. Furthermore, in the context of dopamine depletion in the GP area, a bursting activity in GP that is unique to PD was identified. These findings highlight the significance of considering all pathology-related changes in the network in order to properly investigate its activity. As a result, in neurodegenerative diseases like PD, which induce dopamine depletion and synaptic degenerations, the BG synchronous activity increase and beta rhythm slowdown in the thalamus area may be predicted to occur at the same time. The current study's main focus was the application of DBS to the basal ganglia-thalamus network model and the cerebellar-basal-ganglia thalamocortical network. We used DBS in both states of thecerebellar-basal-ganglia thalamocortical network, at different frequencies, and on four different populations: the thalamus, the GPi, which is a common target for PD, the STN, and the GPe, which are also used as PD targets\cite{yousif2017network,yousif2020population,lu2020erratum}. At low frequencies, DBS reduced the frequency of the oscillations and caused them to occur in bursts \cite{beudel2019linking}. This may be consistent with findings that low-frequency DBS increases pathological activity, especially since pathological neural activity in Parkinson's disease has been shown to contain more bursting modes, for example, in the GPi \cite{yu2019oscillation}. According to our findings, high-frequency DBS delivered to GPi in the striatum can block GPe beta oscillations in the PD state. Moreover, as the frequency of the DBS increases, the beta frequency activity of GPi is suppressed. These conclusions, obtained with an extended model, are consistent with earlier reported results. This research can assist in a better understanding of the role of various brain regions as well as of the processes through which DBS produces a therapeutic impact. Future research should combine this approach with comprehensive single neuron models of the DBS target regions on large scales.

%\backmatter

\section*{Acknowledgments}
Authors are grateful to the NSERC and the CRC Program for their support. RM is also acknowledging support of the BERC 2018-2021 program and Spanish Ministry of Science, Innovation and Universities through the Agencia Estatal de Investigacion (AEI) BCAM Severo Ochoa excellence accreditation SEV-2017-0718, and the Basque Government fund AI in BCAM EXP. 2019/00432.

\subsection*{Author contributions}

HS: methods and materials, data curation, formal analysis, investigation, writing- original draft preparation. RM: conceptualization, supervision and reviews. All authors approved the final submitted version. 

\subsection*{Conflict of interest}

The authors declare no potential conflict of interests.

\nocite{*}% Show all bib entries - both cited and uncited; comment this line to view only cited bib entries;
\bibliography{wileyNJD-AMA}%
\end{document}